# Persistent Ionic Photo-responses and Frank-Condon Mechanism in Proton-transfer Ferroelectrics

Xuanyuan Jiang[1], Xiao Wang[2], Pratyush Buragohain[1], Andy Clark[2], Haidong Lu[1], Shashi Poddar[1], Le Yu[2,3], Anthony D DiChiara[4], Alexei Gruverman[1,5], Xuemei Cheng[2], Xiaoshan Xu[1,5]

[1]Department of Physics and Astronomy, University of Nebraska, Lincoln, NE 68588, USA

[2]Department of Physics, Bryn Mawr College, Bryn Mawr, PA 19010, USA

[3]School of Electronic Science and Engineering, Nanjing University, Nanjing 210093, China

[4]Advanced Photon Source, Argonne National Laboratory, Lemont, IL 60439, USA

[5]Nebraska Center for Materials and Nanoscience, University of Nebraska, Lincoln, NE 68588, USA

Correspondence to: xiaoshan.xu@unl.edu

Abstract

Photoexcitation is well-known to trigger electronic metastable states and lead to phenomena like long-lived photoluminescence and photoconductivity. In contrast, persistent photo-response due to ionic metastable states is rare. In this work, we report persistent structural and ferroelectric photo-responses due to proton metastable states via a nuclear quantum mechanism in ferroelectric croconic acid, in which the proton-transfer origin of ferroelectricity is important for the ionic metastable states. We show that, after photoexcitation, the changes of structural and ferroelectric properties relax in about 1000 s, while the photoconductivity decays within 1 s, indicating the dominant ionic origin of the responses. The photogenerated internal bias field that survives polarization switching process suggests another proton transfer route and metastable state, in addition to the metastable states resulting from proton transfer along the hydrogen bonds proposed previously. Analysis based on Frank Condon principle reveals the quantum mechanical nature of the proton-transfer process both within the hydrogen bonds and out of the hydrogen bonds, where small mass of proton and significant change of potential landscape due to the excited electronic states are the key. The demonstration of persistent photo-responses due to the proton metastable states unveils a nuclear quantum mechanism for photo-tunability of materials, which is expected to impact many material properties sensitive to ionic positions.

Introduction

A material's reversible electronic photo-responses can vary over a large timescale range. For instance, light reflection and absorption are normally due to electronic responses of materials that occur in femtoseconds, while photoexcited electrons may also go into metastable states leading to effects such as hour-long photo luminescence and photo conductivity[1–4]. While the timescale of the electronic photo-responses is clearly a key parameter for applications of optical and transport properties, similar responses due to ionic metastable states, which are rare, is expected to impact applications on most material properties that depends on structures.

The large timescale variation of the electronic photo-responses comes from the direct interaction between light and electrons, which allows the electrons to acquire most of the photon energy to access a large number of states. This is generally not the case for ions. According to the Frank-Condon principle [5–7], the photoexcitation mostly starts with electronic excitation in a "vertical" fashion in femtoseconds without changing the vibrational states of the ions (or nuclei). After that, as illustrated in Figure 1a, ions acquire energy in the form of a few vibrational energy quanta via quantum transition to modified vibrational eigenstates defined by the reshaped potential energy landscape according to the excited electronic states.

Persistent photo-responses require metastable states, which in principle can be found in ferroelectric materials. For example, in ferroelectric $PbTiO_3$, Ti ions sit in asymmetric double-well potentials tilted by the local field, in which the ground state is local in the deeper well and the metastable state may exist in the shallower well [8]. However, the energy of a few vibrational quanta on the order of 10 meV may not be enough to overcome the barrier between the wells to reach the metastable states. As a result, the ionic photo-responses of materials like $PbTiO_3$ is dominated by the classical converse piezoelectric effects following the electronic excitation in the form of photovoltaic effect with a nanosecond timescale [9,10], or simply thermal expansion [11]. If the large vibrational energy quantum is critical to promote the ions to the metastable state, the recently discovered proton-transfer type molecular ferroelectrics [12] are promising candidates for persistent reversible ionic photo-responses [13,14].

A prototypical example of proton-transfer type molecular ferroelectrics is croconic acid (CA), which consists of stacked herringbone layers, [15,16] as shown in Figure 1b. While these layers are held together by the van der Waals interactions, the molecules within a layer are connected by the (ridge and plane) hydrogen bonds (H-bond). As depicted in Figure 1c, the protons in the hydrogen bonds have two energy minima resulting from the double-well potentials. CA crystals exhibit spontaneous polarization due to the ordering of proton positions on the same side of the double wells with a Curie temperature higher than 400 K [16]. Polarization switching corresponds to the collective transfer of protons to the other wells, or transition between two tautomers [15]. The vibrational energy quantum can be estimated as ~ 0.1 eV using uncertainty principle for a proton in a well of about 0.5 Å wide which is comparable to the hydrogen-bond energy [17]; the large energy quantum is consistent with the fact that proton positions remain ordered in CA solid phase [18]. Therefore, the quantum transition of vibrational states resulting from the Frank Condon principle is potentially a viable process to transfer protons to the metastable stats in the other wells. In addition, croconic acid has a large spontaneous polarization ($\approx$ 20 µC/cm$^2$) [15,16] originating from both the ordered protons and the distorted electronic cloud [13,19–21]. The electronic excitation is expected to have a significant impact on the proton potential well [13,22].

In this work, we studied the responses of structural, ferroelectric, and electric transport properties to photoexcitation in CA films. Reversible changes have been observed in crystal structure and piezoresponse on the order of 1000 s, but not in electric conductivity, indicating dominant ionic contributions. The observed persistent internal bias field that survives polarization switching suggests additional proton transfer route out of the hydrogen bonds.

Results

The ionic metastable states are expected to perturb the crystal structure. More specifically, lattice expansion is expected for CA crystals after photoexcitation, because it reduces the order of the positions of protons and molecules which are otherwise tightly packed. We therefore studied the photostriction and its timescale of polycrystalline CA films using the time-resolved x-ray diffraction (XRD), as demonstrated previously on other ferroelectric materials [11,10,23]. The experimental setup of time-resolved XRD is depicted in Figure 2a. The film samples were excited using a 330 nm (3.8 eV, above the 3.1 eV band gap [24]) laser; a low fluence of 100 mW/cm$^2$ was chosen to avoid sample damage and irreversible structural changes. The crystal structure of croconic acid film were monitored using powder x-ray diffraction. The lattice constants were extracted from the structural refinement using the software Fullprof based on the Pca2$_1$ crystal structure (Supplementary Information, Section 1) [16,25–27]; the results are plotted in Figure 2b and 2c as a function of time.

Starting at zero time in Figure 2b, the film sample was photoexcited. All lattice constants increase, consistent with the expected positive photostriction. At about 300 s, while the lattice expansion continues, the rate of lattice expansion reduces; this process remains for the rest of the measurement with the photoexcitation. Figure 2c displays the lattice change after the photoexcitation is stopped at zero time. The lattice constants relax back in ~1000 s to approximately the original values before the photoexcitation.

To check the impact of temperature increase (due to the light absorption) to the lattice expansion, we measured the thermal expansion of the CA films directly. As shown in Figure 2d, when the CA films were heated, all lattice constants increase compared with the room-temperature values, but with a large anisotropy: the thermal expansion coefficients are 50, 80, and 27 ppm/K along the *a*, *b*, and *c* axes, respectively. The anisotropy of the thermal expansion is consistent with the crystal structural anisotropy of CA [16,27] and previous measurements [28]: the largest expansion is along the *b* axis which is the direction the herringbone layers are held together by the van der Waals interaction. The large difference in anisotropy between Figure 2b and Figure 2d suggests that the impact of thermal expansion to the observed photostriction is less important. Therefore, the observed persistent photostriction is more likely from the metastable states.

The ionic metastable states may also manifest in the ferroelectric properties of the CA films. Particularly, the proton displacements generate local electric dipole moments. To reveal the photoinduced modification of polarization, we studied the ferroelectric switching behavior of the photoexcited CA films using piezoresponse force microscopy (PFM) [29–31]. Previously, we demonstrated that continuous CA films consisting of nanometer-sized grains can be prepared on conducting oxide films [32–34] and single-grain ferroelectric switching can be achieved using the conductive tip of a scanning probe as the top electrode [35]. As depicted in Figure 3a, using the PFM tip, the time-evolution of piezoresponse of CA grains were measured. Clear butterfly-type hysteresis loops have been observed, indicative of polarization switchability. The coercive voltage is about 6 V (Supplementary Information, Section 2, Fig. S2a), in agreement with our previous results measured on films grown under similar conditions [35].

Interestingly, the piezoresponse hysteresis loops, after photoexcitation (390 nm, 3.2 eV photon, 15 mW/cm$^2$), show dramatic asymmetry in Figure 3b. The hysteresis loops are highly shifted to the right immediately after stopping the photoexcitation; correspondingly, the piezoresponses for the positively and negatively poled states at zero bias show large contrasts (> 2). The voltage shift of the hysteresis loop in Figure 3b indicates the existence of an internal bias field. The asymmetry slowly decays overtime after the photoexcitation is turned off and the hysteresis loop becomes symmetric after about 1 hour. To analyze the time evolution of the asymmetry, we plotted the internal bias as a function of time after the photoexcitation in Figure 3c; a fit assuming exponential decay results in a time constant of $1.5 \times 10^3$ s, consistent with the timescale observed in photostriction.

The reversible persistent photostriction and photo-induced changes in piezoresponses, suggest that CA samples enter the metastable states after the photoexcitation. Our photoconductivity measurements (Supplementary Information, Section S3), which reveals a < 1 s relaxation time, corroborates that these metastable states are of ionic origin. More specifically, the magnitude of the photostriction and internal bias are consistent with the proton-transfer type metastable states (Supplementary Information, Section S2).

In proton-transfer type ferroelectrics, polarization switching corresponds to protons transfer between two equivalent positions in double-well potentials collectively. For a certain molecule in a poled ferroelectric material, the double-well is tilted by the local electric field, generating a stable state and a metastable state. Previous studies of proton-transfer type ferroelectric crystals using optical second harmonic generation indicated a component of the photo-responses with a long timescale [13,14]. The observation was attributed to the metastable states resulting from proton transfer along the hydrogen bonds [13,14,22], indicated as route A in Figure 1c. The configuration can be viewed as the reversed single-molecule ferroelectric domain. We call this domain-like metastable state.

Besides the domain-like metastable state, the ~1000 s timescale observed in both structural and ferroelectric responses, suggests an additional metastable state. The longer timescale of this metastable state is consistent with a larger energy scale than that of the domain walls since the polarization switching process cannot remove this metastable state. We propose that this metastable state results from proton transfer to another nearby oxygen site, indicated as route B in Figure 1c. Also depicted in Figure 1c is the slight bending (155 degree) of hydrogen bond toward the other oxygen along the route B in the crystal structure [36], suggesting sizable attractive force. In this metastable state, with the broken hydrogen bond, the hydrogen is bonded to only one oxygen site, generating an interstitial/vacancy defect. We call this defect-like metastable state. The proton is further away from the originally bonded oxygen sites, making it more difficult to return to the ground state, in line with the internal bias field that cannot be removed by the polarization switching process. We note that the possibility that the charge carriers in trap states[37] lead to the internal bias can be ruled out according to our photoconductivity measurements (Supplementary Information, Section S3).

Next, we investigate the possibility of these metastable states and the corresponding proton-transfer routes. Previous calculations found that the lowest excited electronic state of the CA molecules carries a dipole moment opposite to the polarization direction of the CA crystal.[13,22] Based on these findings, simulations assuming adiabatic processes suggest that the photo-induced electronic excitation causes proton transfer to the other side of the hydrogen bonds,[13] generating metastable states in the form of local reversed ferroelectric domains.

To account for both the domain-like metastable states and the defect-like metastable states, we analyze the proton-transfer process in terms of the response of proton vibrational states to the change of double well after the "vertical" electronic excitation. We employ a model in which most parameters can be estimated using the known CA crystal structure and the hydrogen-bond energy. The simplicity of the model allows a clear illustration of the physical process and the investigation of multiple proton transfer routes.

The potential energy of a proton is assumed to come from two neighboring oxygen sites separated by distance $D_0$:

$$V_{DW}(x) = -\frac{a(1-\delta)}{\left(x-\frac{D_0}{2}\right)^4} + \frac{b}{\left(x-\frac{D_0}{2}\right)^8} - \frac{a(1+\delta)}{\left(x+\frac{D_0}{2}\right)^4} + \frac{b}{\left(x+\frac{D_0}{2}\right)^8}.$$

where $a$ and $b$ are parameters for the attractive and repulsive forces of the oxygen sites, $\delta$ is the asymmetry parameter. For CA crystals, $a$ and $b$ can be determined using hydrogen-bond energy $E_{HB}$, proton-oxygen distance $x_0$, and the hydrogen-bond length $L$, which are known from the CA crystal structure (see Supplementary Information Section 4). The variable $\delta$ describes the change of the double-well potential caused by the electronic excitation. One can numerically solve the vibrational eigenstates by diagonalizing the Hamiltonian [38]

$$H = -\frac{\hbar^2}{2m}\frac{\partial^2}{\partial x^2} + V_{DW}(x).$$

To study the proton transfer along route A (for domain-like metastable states), we simulated the vibrational states of protons using the structural parameters $x_0 = 0.977$ Å, $L = 1.641$ Å, and $D_0 = 2.618$ Å (Supplementary Information Section 4) and assuming $E_{HB} = 0.3$ eV.[17,39] The results are displayed in Figure 1a.

For the electronic ground state, the double-well potential has a significant asymmetry described by $\delta_g = 0.1$, due to the local field generated by the spontaneous polarization. The energy barrier in the middle is consistent with the result from the previous first-principles calculations.[13] The vibrational ground state $\psi_0$ is local in the left well with an energy lower than the barrier. The ~ 0.1 eV separation between the $\psi_0$ and $\psi_1$ is consistent with the estimated vibrational energy quantum from the uncertainty principle, which means that the protons stay mostly in the ground state $\psi_0$.

For the electronic excited state of the CA molecule, since it carries a dipole moment [13,22] that generates a field opposite to the local field and reduces the asymmetry of the double well, we represent the effect using a smaller value $\delta_e = 0.01$. The two local states, $\psi_0^e$ and $\psi_1^e$, due to the reduced asymmetry, change substantially from $\psi_0$ and $\psi_1$ respectively, suggesting a sizable probability for the proton to transfer from $\psi_0$ (left well) to $\psi_1^e$ (right well). Figure 4a displays the transition probability $T_n = |\langle \psi_n^e | \psi_0 \rangle|^2$ as a function of the vibrational energy of the $\psi_n^e$ state. The probability drops quickly with energy because the higher energy states are less affected by the asymmetry change of the well and tend to remain orthogonal to $\psi_0$. Therefore, the proton can only acquire efficiently a few vibrational energy quanta.

The change of asymmetry from $\delta_g$ to $\delta_e$ is critical for the proton transfer. The total probability of proton transfer from the left well to the right well is $T_{pt} = \sum T_i = \sum |\langle \psi_i^e | \psi_0 \rangle|^2$ for all $\psi_i^e$ states localized in the right well. For the ionic potential in Figure 1a, $T_{pt} = T_1$ since $\psi_1^e$ is the only local state in the right well. As shown in Figure S6, $T_1$ is calculated as a function of $\delta_g$ and the $\delta_e/\delta_g$ ratio. When $\delta_e/\delta_g \to 1$, i.e., toward no change of asymmetry, $T_{pt}$ drops quickly. This is expected because when there is no change of asymmetry, $\psi_n^e$ remains orthogonal to $\psi_0$. In contrast, $T_{pt}$ is much less sensitive to $\delta_g$. The insensitivity of $T_{pt}$ to $\delta_g$ makes the model more robust.

To study the proton transfer along route B (for defect-like metastable states), we calculated the vibrational states of protons using a longer O-O distance $D_0 = 3.18$ Å in the model (Supplementary Information Section 4), with other parameters kept the same. The results are displayed in Figure 4b. As the two wells are more separated, the states are more localized compared with those in Figure 1a. In addition, larger $D_0$ also makes the barrier between the two wells higher for there is less attraction from the other oxygen site. As a result, additional local states appear. Figure 4c shows the calculated $T_{pt}$ as a function of $D_0$. In general, larger $D_0$ reduces $T_{pt}$ due to the localization of the vibrational states. On the other hand, when additional local states appear, $T_{pt}$ gets boosted. This can be an important mechanism for protons to follow the route B to reach the defect-like metastable state.

Overall, the double-well model offers a plausible quantum mechanical mechanism to account for the experimental observation with the domain-like metastable state with smaller timescale and defect-like metastable state with a larger timescale. The key factors for protons to reach these metastable states are (1) A significant change of double-well symmetry that modifies the vibrational eigenstates. This comes from the dipole moment of the electronic excited state of the CA molecules. (2) Few local vibrational eigenstates so that the change of double-well symmetry can efficiently modify these states. This is enabled by the small mass of proton which makes the vibrational energy quantum comparable to the barrier between the two wells.

## Conclusion

We have demonstrated persistent reversible photostriction effect (~ 1000 s) in croconic acid using time-resolved x-ray diffraction. The effect correlates with the buildup of the internal electric field which relaxes in a similar time scale, as indicated by local PFM spectroscopy measurements, while photoconductivity decay within 1 s. These observations indicate an ionic origin of these persistent photo-responses related to metastable states of protons due to photoexcitation. Analysis based on Frank Condon principle reveals that the small mass of proton and the change of symmetry in the potential energy of proton are key factors for the protons to transfer to the metastable states via quantum transition, which can occur both along the hydrogen bonds and out of the hydrogen bonds. These results uncover the persistent reversible ionic responses to photoexcitation in proton-transfer type molecular ferroelectrics with a nuclear quantum mechanism. The metastable proton states in turn impact the material properties like crystal structure, ferroelectricity, and potentially many other properties sensitive to crystal structure, which is promising for diverse applications.

## Experimental Methods

Croconic acid (CA) films have been fabricated by physical vapor deposition (PVD) in high vacuum ($1\times10^{-7}$ Torr) with an EvoVac system from Angstrom Engineering on different substrates: 300 nm films on $Al_2O_3$ substrate for time-resolved x-ray diffraction (XRD), 140 nm films on patterned $Au/SiO_2$ substrate with 60 nm $Al_2O_3$ capping layer for photocurrent measurements, and 40 nm films on conductive $NiCo_2O_4$ (NCO)/$MgAl_2O_4$ for piezoresponse force microscopy (PFM). The CA films for the PFM studies were grown with a -33 °C substrate temperature for smooth surface morphology followed by postgrowth annealing at 30 °C for 1 hour to ensure crystallization; all the other CA films were grown at room temperature.

The time-resolved x-ray diffraction on CA films was carried out at room temperature at beam line 14-ID-B at the Advanced Photon Source in the Argonne National Lab. The samples were placed 32 cm away from a two-dimensional detector. Optical excitation during the x-ray diffraction studies was provided by laser pulses (500 Hz) with 2 μj/mm$^2$ (100 mW/cm$^2$) fluence and 330 nm wavelength (3.8 eV) with an electronically adjustable time delay. Incident x-ray pulses with a photon energy 12 keV were focused to an area of 75 $\mu$m full-width half-maximum (FWHM) in diameter with a 1.5 deg incident angle, overfilled by a 120 $\mu$m × 3400 $\mu$m pump laser spot.

The AC photocurrent was measured by an impedance analyzer Solartron 1260, with 1 V voltage and 100 Hz frequency. The dimension of the conduction channel is 10 $\mu m$ × 100 $nm$ × 80 $nm$. The light source is a commercial diode laser with 400 nm wavelength (3.1 eV) with 140 mW/cm$^2$ fluence.

Local PFM spectroscopic studies were carried out by means of a commercial AFM system (MFP-3D, Asylum Research) using Cr/Pt/Ir-coated Si probes (PPP-EFM, NANOSENSORS). All PFM measurements were done in the resonant enhanced mode using a ~350 kHz AC signal with 0.8 V drive amplitude. The bias was applied to the tip and the bottom electrode was grounded in the PFM measurements. The light source used in the PFM study has 390 nm wavelength (3.2 eV) with 15 mW/cm$^2$ fluence.

## Acknowledgement

This research was primarily supported by the U.S. Department of Energy (DOE), Office of Science, Basic Energy Sciences (BES), under Award No. DE-SC0019173 (device fabrication, transport measurements, x-ray diffraction, and modeling). Additional support was from the National Science Foundation (NSF) under Grant No. DMR-1420645 (PFM). Work at Bryn Mawr College was supported by NSF DMR-1708790. This research used resources of the Advanced Photon Source, a U.S. Department of Energy (DOE) Office of Science User Facility operated for the DOE Office of Science by Argonne National Laboratory under Contract No. DE-AC02-06CH11357. Use of BioCARS was also supported by the National Institute of General Medical Sciences of the National Institutes of Health under Grant No. R24GM111072. The content is solely the responsibility of the authors and does not necessarily represent the official views of the National Institutes of Health (NIH). Time-resolved setup at Sector 14 was funded in part through a collaboration with Philip Anfinrud (NIH/National Institute of Diabetes and Digestive and Kidney Diseases). The research was performed in part in the Nebraska Nanoscale Facility: National Nanotechnology Coordinated Infrastructure and the Nebraska Center for Materials and Nanoscience (and/or NERCF), which are supported by the National Science Foundation under Award ECCS: 2025298, and the Nebraska Research Initiative.

Data availability

The datasets generated during and/or analyzed during the current study are available from the corresponding author on reasonable request.

Author Contributions

The thin films were synthesized by X.J. with the assistance from X.X. Time-resolved crystal structure was characterized by X.J., X.W., A. C. and L. Y. with the help of A.D. The time-resolved ferroelectric properties were characterized by P.B. and H.L, under the supervision of A.G. The photocurrent measurements were characterized by X.J. X.X. carried out the modelling to explain the observations. X.J. and X.X. wrote the manuscript. The study was conceived and guided by X.X. All authors discussed results and commented on the manuscript.

Author Information

The authors declare no competing financial interests. Readers are welcome to comment on the online version of the paper. Correspondence and requests for materials should be addressed to X.X. (xiaoshan.xu@unl.edu).

**References**

1. Matsuzawa, T., Aoki, Y., Takeuchi, N. & Murayama, Y. A New Long Phosphorescent Phosphor with High Brightness, SrAl2O4:Eu2+,Dy3+. *J. Electrochem. Soc.* **143**, 2670 (1996).

2. Tarun, M. C., Selim, F. A. & McCluskey, M. D. Persistent photoconductivity in strontium titanate. *Phys. Rev. Lett.* **111**, 187403 (2013).

3. Yin, H., Akey, A. & Jaramillo, R. Large and persistent photoconductivity due to hole-hole correlation in CdS. *Phys. Rev. Mater.* **2**, 084602 (2018).

4. Kabe, R. & Adachi, C. Organic long persistent luminescence. *Nature* **550**, 384–387 (2017).

5. Condon, E. A theory of intensity distribution in band systems. *Phys. Rev.* **28**, 1182–1201 (1926).

6. Franck, J. & Dymond, E. G. Elementary processes of photochemical reactions. *Trans. Faraday Soc.* **21**, 536–542 (1926).

7. Condon, E. U. Nuclear motions associated with electron transitions in diatomic molecules. *Phys. Rev.* **32**, 858–872 (1928).

8. Joseph, J., Vimala, T. M., Sivasubramanian, V. & Murthy, V. R. K. Structural investigations on Pb(ZrxT1−x)O3 solid solutions using the X-ray Rietveld method. *J. Mater. Sci.* **35**, 1571–1575 (2000).

9. Matzen, S. *et al.* Tuning Ultrafast Photoinduced Strain in Ferroelectric-Based Devices. *Adv. Electron. Mater.* **5**, 1800709 (2019).

10. Daranciang, D. *et al.* Ultrafast Photovoltaic Response in Ferroelectric Nanolayers. *Phys. Rev. Lett.* **108**, 087601 (2012).

11. Sinha, K. *et al.* Effects of biaxial strain on the improper multiferroicity in h-LuFeO$_3$ films studied using the restrained thermal expansion method. *Phys. Rev. B* **95**, 094110 (2017).

12. Horiuchi, S., Kumai, R. & Tokura, Y. Hydrogen-bonding molecular chains for high-temperature ferroelectricity. *Adv. Mater.* **23**, 2098–2103 (2011).

13. Iwano, K. *et al.* Ultrafast Photoinduced Electric-Polarization Switching in a Hydrogen-Bonded Ferroelectric Crystal. *Phys. Rev. Lett.* **118**, 107404 (2017).

14. Umanodan, T. *et al.* Different time-scale relaxation dynamics in organic supramolecular ferroelectrics studied by linear and nonlinear spectroscopy. *J. Phys. Soc. Japan* **84**, 073707 (2015).

15. Horiuchi, S., Kobayashi, K., Kumai, R. & Ishibashi, S. Proton tautomerism for strong

polarization switching. *Nat. Commun.* **8**, 14426 (2017).

16. Horiuchi, S. *et al.* Above-room-temperature ferroelectricity in a single-component molecular crystal. *Nature* **463**, 789–792 (2010).

17. Wendler, K., Thar, J., Zahn, S. & Kirchner, B. Estimating the Hydrogen Bond Energy. *J. Phys. Chem. A* **114**, 9529–9536 (2010).

18. Mukhopadhyay, S. *et al.* Mechanism of enhancement of ferroelectricity of croconic acid with temperature. *Phys. Chem. Chem. Phys.* **19**, 32216–32225 (2017).

19. Tang, F. *et al.* Probing ferroelectricity by X-ray absorption spectroscopy in molecular crystals. *Phys. Rev. Mater.* **4**, 034401 (2020).

20. Cai, Y., Luo, S., Zhu, Z. & Gu, H. Ferroelectric mechanism of croconic acid: A first-principles and Monte Carlo study. *J. Chem. Phys.* **139**, 044702 (2013).

21. Seliger, J., Plavec, J., Sket, P., Zagar, V. & Blinc, R. 17O NQR and 13C NMR study of hydrogen-bonded organic ferroelectric croconic acid. *Phys. status solidi B* **248**, 2091–2096 (2011).

22. Miyamoto, T. *et al.* Ultrafast polarization control by terahertz fields via π-electron wavefunction changes in hydrogen-bonded molecular ferroelectrics. *Sci. Rep.* **8**, 15014 (2018).

23. Wen, H. *et al.* Electronic origin of ultrafast photoinduced strain in BiFeO3. *Phys. Rev. Lett.* **110**, 037601 (2013).

24. Sawada, R. *et al.* Large second-order optical nonlinearity in a ferroelectric molecular crystal of croconic acid with strong intermolecular hydrogen bonds. *Appl. Phys. L* **102**, 162901 (2013).

25. Bisti, F. *et al.* The electronic structure of gas phase croconic acid compared to the condensed phase: More insight into the hydrogen bond interaction. *J. Chem. Phys.* **138**, 014308 (2013).

26. Bisti, F., Stroppa, A., Picozzi, S. & Ottaviano, L. Fingerprints of the hydrogen bond in the photoemission spectra of croconic acid condensed phase: an x-ray photoelectron spectroscopy and ab-initio study. *J. Chem. Phys.* **134**, 174505 (2011).

27. Di Sante, D., Stroppa, A. & Picozzi, S. Structural, electronic and ferroelectric properties of croconic acid crystal: a DFT study. *Phys. Chem. Chem. Phys.* **14**, 14673 (2012).

28. Mukhopadhyay, S., Gutmann, M. & Fernandez-Alonso, F. Hydrogen-bond structure and anharmonicity in croconic acid. *Phys. Chem. Chem. Phys.* **16**, 26234–26239 (2014).

29. Gruverman, A., Auciello, O. & Tokumoto, H. Imaging and Control of Domain Structures in Ferroelectric Thin Films Via Scanning Force Microscopy. *Annu. Rev. Mater. Sci.* **28**,

101–123 (1998).

30. Gruverman, A. & Kalinin, S. V. Piezoresponse force microscopy and recent advances in nanoscale studies of ferroelectrics. *J. Mater. Sci.* **41**, 107–116 (2006).

31. Bonnell, D. A., Kalinin, S. V., Kholkin, A. L. & Gruverman, A. Piezoresponse Force Microscopy: A Window into Electromechanical Behavior at the Nanoscale. *MRS Bull.* **34**, 648–657 (2009).

32. Yuan, Y., Jiang, X., Poddar, S. & Xu, X. Electric-field assisted nucleation processes of croconic acid films. *CrystEngComm* **21**, 7460–7467 (2019).

33. Zhen, C. *et al.* Nanostructural origin of semiconductivity and large magnetoresistance in epitaxial NiCo2O4/Al2O3 thin films. *J. Phys. D. Appl. Phys.* **51**, 145308 (2018).

34. Mellinger, C., Waybright, J., Zhang, X., Schmidt, C. & Xu, X. Perpendicular magnetic anisotropy in conducting NiCo2 O4 films from spin-lattice coupling. *Phys. Rev. B* **101**, 14413 (2020).

35. Jiang, X. *et al.* Room temperature ferroelectricity in continuous croconic acid thin films. *Appl. Phys. Lett.* **109**, 102902 (2016).

36. Braga, D., Maini, L. & Grepioni, F. Crystallization from hydrochloric acid affords the solid-state structure of croconic acid (175 years after its discovery) and a novel hydrogen-bonded network. *CrystEngComm* **3**, 27 (2001).

37. Dimos, D., Warren, W. L., Sinclair, M. B., Tuttle, B. A. & Schwartz, R. W. Photoinduced hysteresis changes and optical storage in (Pb,La)(Zr,Ti)O 3 thin films and ceramics. *J. Appl. Phys.* **76**, 4305–4315 (1994).

38. Marsiglio, F. The harmonic oscillator in quantum mechanics: A third way. *Am. J. Phys.* **77**, 253–258 (2009).

39. Fernandez-Alonso, F. *et al.* Hydrogen Bonding in the Organic Ferroelectric Croconic Acid: Insights from Experiment and First-Principles Modelling. *J. Phys. Soc. Japan* **82**, SA001 (2013).

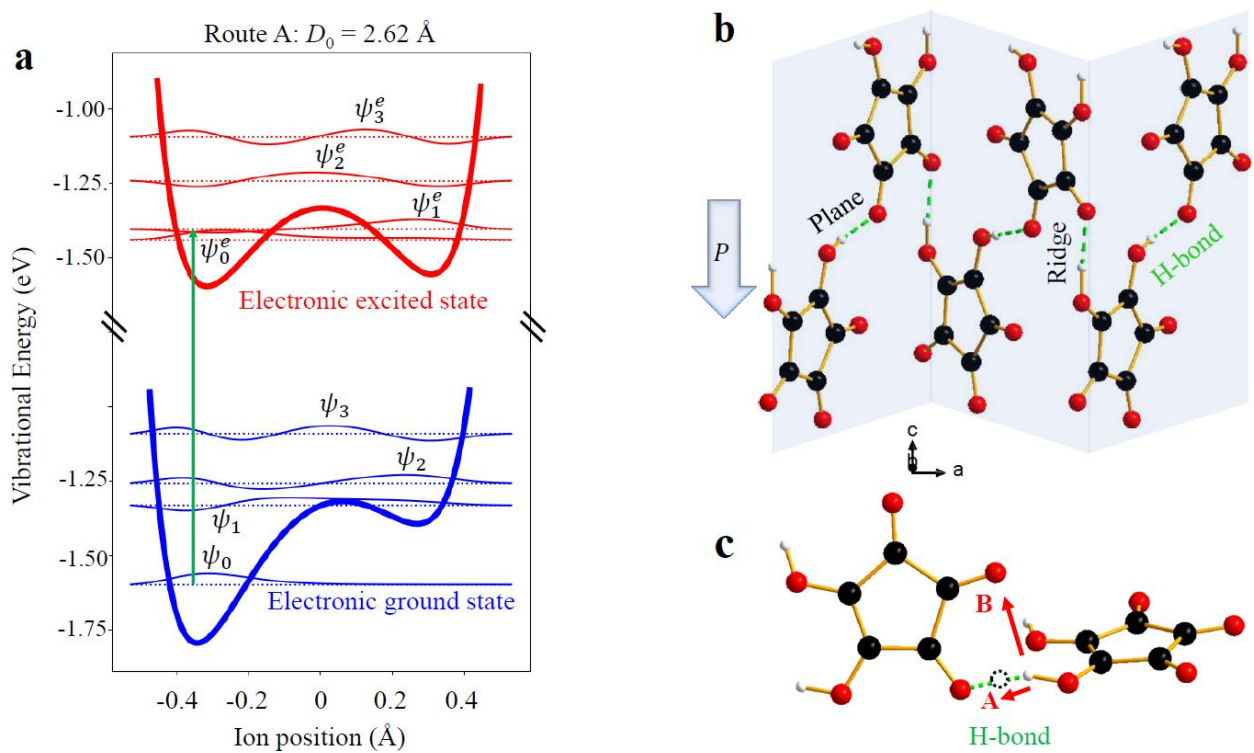

**Figure 1.** Illustration of proton transfer process and the structure of croconic acid. (a) Simulated potential energy landscapes (thick lines) and the vibrational eigenstates (thin lines) of protons in croconic acid hydrogen bonds with the electronic ground state and the excited state respectively. The eigenstate wave functions are plotted on the dashed baselines indicating the energies. The vertical arrow connects the initial and final states for the possible proton transfer after the "vertical" electronic excitation. (b) Fragment of a herring-bone layer of the croconic acid crystal structure. The spontaneous polarization is pointing down due to the ordering of proton positions in the intermolecular hydrogen bonds along the dashed line. (c) Two croconic acid molecules connected by a ridge hydrogen bond. The dashed circle indicates the other energy minimum position of the proton. The arrows indicate the two potential proton transfer routes.

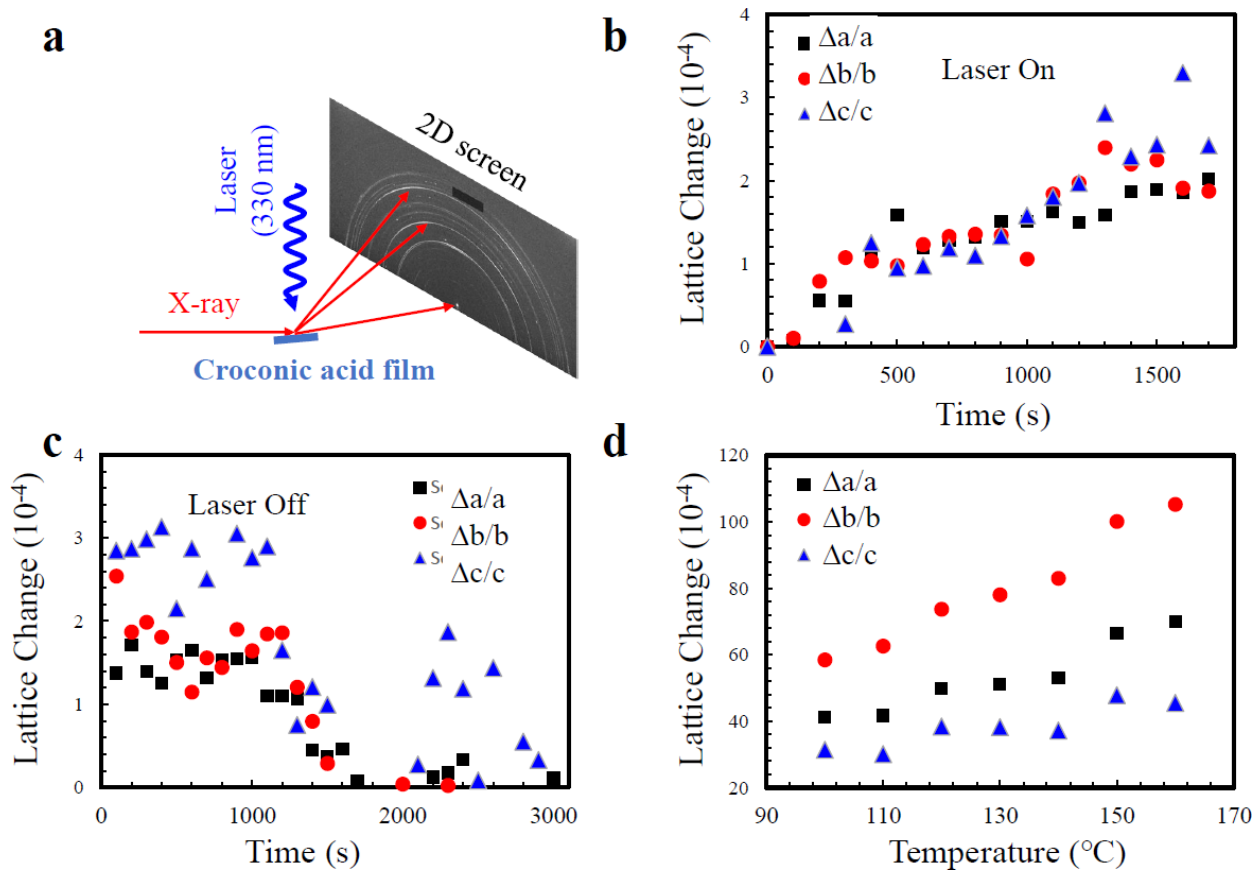

**Figure 2.** Photostriction in croconic acid thin films and its relaxation. (a) Experimental setup for time-resolved synchrotron x-ray diffraction. (b) Time-resolved lattice distortion along three axes under photoexcitation starting at zero time. (c) Lattice relaxation after the laser is turned off at zero time. (d) Lattice change due to the thermal expansion with a large anisotropy with respect to the room-temperature values.

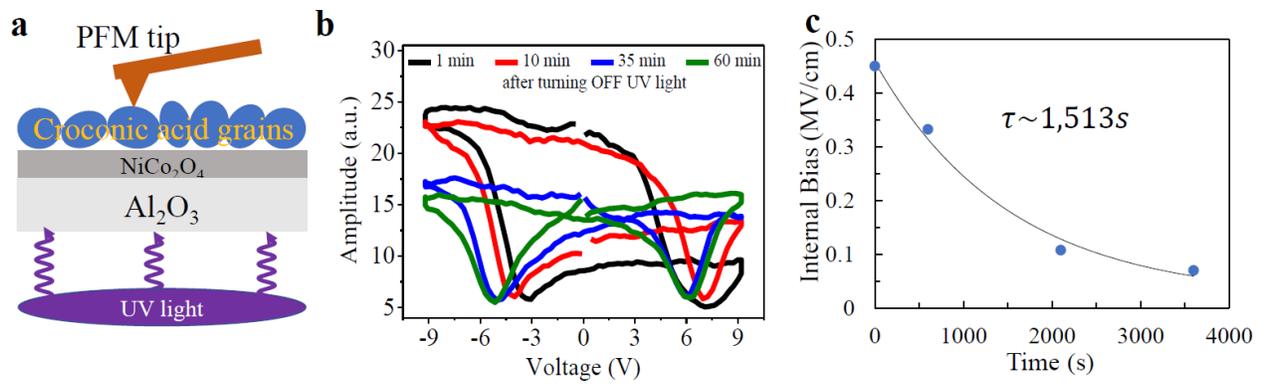

**Figure 3**. Effect of photoexcitation on the piezoresponse of croconic acid grains. (a) Schematic diagram of the internal bias study via piezoresponse force microscope (PFM) under the illumination. (b) Hysteretic piezoresponse amplitude at different time after photoexcitation stops, showing shape distortions and shifts of the response along the voltage axis. (c) Internal bias from (b) as a function of relaxation time and the corresponding fit assuming exponential decay. The error bar is 0.0125 MV/cm.

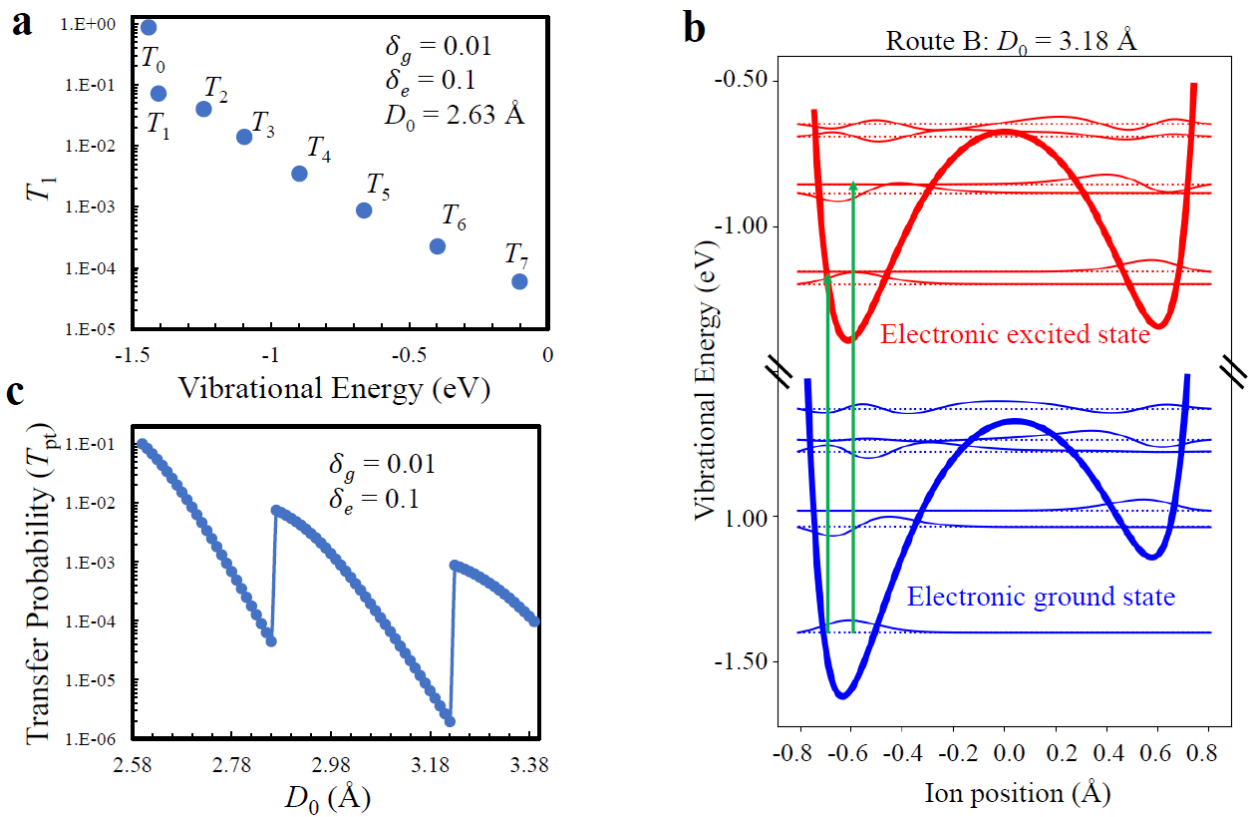

**Figure 4**. (a) Transition probability from the vibrational ground state to the modified eigenstates due to the change of potential energy landscape. (b) Simulated potential energy landscapes (thick lines) and the vibrational eigenstates (thin lines) of protons with the electronic ground state and the excited state respectively. The eigenstate wave functions are plotted on the dashed baselines indicating the energies. The vertical arrows connect the initial and final states for the potential proton transfer after the "vertical" electronic excitation. (c) The proton transfer probability as a function of oxygen-oxygen site distance.

15